\documentstyle[12pt,aaspp4,flushrt]{article}

\newcommand{\bq}{\begin{equation}}
\newcommand{\eq}{\end{equation}}
\newcommand{\mach}{{\mathcal M}}
\newcommand{\kpc}{\mbox{ kpc}}
\newcommand{\mpc}{\mbox{ Mpc}}
\newcommand{\hunits}{\mbox{ km s$^{-1}$ Mpc$^{-1}$}}

\newcommand{\cmden}{\mbox{ cm$^{-3}$}}
\newcommand{\kms}{\mbox{ km s$^{-1}$}}

\newcommand{\sigm}{\mbox{ cm$^2$ g$^{-1}$}}
\newcommand{\pc}{\mbox{ pc}}

\begin{document}
\vskip 0.6 in
\noindent
 
\title{Constraining the Collisional Nature of the Dark Matter\\Through
Observations of Gravitational Wakes}

\author{Steven R. Furlanetto \& Abraham Loeb}
\affil{Harvard-Smithsonian Center for Astrophysics, 60 Garden St.,
Cambridge, MA 02138;\\sfurlanetto@cfa.harvard.edu, aloeb@cfa.harvard.edu}

\begin{abstract}

We propose to use gravitational wakes as a direct observational probe of
the collisional nature of the dark matter.  We calculate analytically the
structure of a wake generated by the motion of a galaxy in the core of an
X-ray cluster for dark matter in the highly-collisional and collisionless
limits.
We show that the difference between these limits can be recovered from
detailed X-ray or weak lensing observations.  We also discuss the sizes of
sub-halos in these limits.  Preliminary X-ray data on the motion of NGC
1404 through the Fornax group disfavors fluid-like dark matter but does not
exclude scenarios in which the dark matter is weakly collisional.

\end{abstract}


\section{Introduction}

Although the dark matter is known to dominate the mass budgets of bound
systems such as galaxies or galaxy clusters, its nature remains
mysterious.  On cosmological scales ($\gg 1 \mpc$), the standard ``cold
dark matter'' (CDM) model, in which the dark matter is non-baryonic,
non-relativistic, and collisionless, has had considerable success matching
observations.  However, two sets of problems have emerged in applying the
same model to smaller scales (\cite{spergel}; \cite{hogan} and references
therein).  The first is known as the ``substructure problem.''  The CDM
model predicts significantly more substructure (in the form of relic dark
matter halos) on small scales than is observed.  For example, models of
the formation of the Local Group of galaxies predict an order of magnitude
more dwarf halos than the observed number of dwarf galaxies
(\cite{moore-lg}).  The second problem is known as the ``cusp problem'':
with no interactions to prevent collapse, CDM halos should form nearly
singular cores (e.g., \cite{nfw}).  In contrast, observations indicate that
galaxies have flat dark matter profiles near their centers (\cite{flores};
Borriello \& Salucci 2001).

These problems led Spergel \& Steinhardt (2000) to reintroduce
self-interacting dark matter (SIDM), originally proposed by
Carlson, Machacek, \& Hall (1992; see also \cite{delaix}).  In this
scheme, the dark matter 
has a significant cross-section for self-interaction (and possibly a
significant cross-section for interaction with baryons as well; see
\cite{qball}).  In such a picture, cusp formation is halted by 
heat conduction (which heats up the dark matter at the halo centers)
and the formation of shocks.  The substructure problem is avoided
because dark matter particle interactions evaporate small halos on
relatively short timescales. 

Existing limits from laboratory experiments restrict the range of
cross-sections available to SIDM that interacts directly with baryons
(\cite{mcguire}) but place no restrictions on the cross-section for dark
matter-dark matter interactions.  To constrain this cross-section, we
must turn to astronomical observations.  In this paper, we propose such
a test, based on observations of the motion of individual galaxies inside
galaxy clusters.

As a galaxy moves through a cluster, the gravitational potential of the
galaxy induces a wake in the surrounding medium.  The back-reaction of this
wake on the galaxy is called dynamical friction and is known to be
important in the dynamics of cluster galaxies (as well as satellite
galaxies and globular clusters on smaller scales; \cite{binney},
\cite{frenk}).  Wakes are induced in both the dark and gaseous components
of the intracluster medium (ICM).  The structure of the dark matter wake
provides a signature of the collisional nature of the dark matter that is
potentially observable through its gravitational lensing effect.  The total
gravitational potential also imprints a wake in the gas distribution; a
precise measurement of the morphology of this wake yields the size of the
perturber's dark matter halo.  This in turn constrains the collisional
nature of the dark matter: the sizes of CDM halos are set by tidal
stripping in the cluster potential (\cite{merritt}), while ram pressure
stripping truncates SIDM halos at considerably smaller radii.  The
exquisite resolution of the \emph{Chandra} satellite allows very precise
measurements of the extent of surface brightness variations in the ICM gas
and could potentially constrain the collisional properties of the dark
matter.

In order to approach this problem analytically, we will focus on two
extreme scenarios for the nature of the dark matter.  The first is the CDM
model, in which the dark matter is collisionless.  The second model assumes
that the dark matter is strongly self-interacting, so that it may be
treated as a perfect fluid.  We will refer to this scenario as the Fluid
Dark Matter (FDM) model.  The latter approximation has been used before to
isolate the implications of SIDM models (\cite{yoshida};
\cite{moore-coll}).  Our calculations will thus bracket the parameter space
available for dark matter self-interaction cross-sections.  
Our experimental test is easiest to interpret in the limit where the
perturber mass is much smaller than that of the host cluster, and so we use
linear theory in our FDM calculations.

We first outline our formalism for calculating the wake structure induced
by a perturber moving through collisionless and collisional media in \S 2.
We then describe the effects of the ICM dark matter on the perturber halo
in \S 3.  Finally, we present our results and compare them to simulations
in \S 4, and we discuss their implications in \S 5.

\section{Gravitational Wakes Due to Dynamical Friction}

First, we develop the mathematical formalism necessary to calculate the
density disturbance in a cluster due to the gravitational effect of a
moving galaxy or group of galaxies.  In \S 2.1 and \S 2.2, we calculate the
wake in collisionless and collisional media, respectively.  In a
collisional medium, an additional wake will be created through deflection
of the ambient medium around the surface of the perturber.  We discuss this
``hydrodynamic wake'' in \S 2.3.  We assume throughout that the perturber
(e.g., a galaxy) with mass $M_g$ moves through a stationary, uniform
background medium (e.g., a cluster core) of density $\rho_c$ with a
constant velocity $v_g = \mach \sigma_c$ directed along the $z$-axis, where
$\mach$ is the Mach number of the flow and $\sigma_c$ is the velocity
dispersion of the background medium (assumed to be equal to the sound speed
for simplicity).  We include a discussion of when the requirement of a
uniform background density may be relaxed in \S 4.

\subsection{ Collisionless Medium }

First we consider the structure of the wake induced in a collisionless
medium, following the method first introduced by Mulder (1983; see also
\cite{weinberg}).  Without collisions, the particle distribution function
$f({\bf x}, {\bf v})$ must be a function only of the integrals of motion
(\cite{binney}), which can be calculated for particles deflected by the
perturber using planar orbit theory (Mulder 1983).  The precise form of the
distribution function is then determined by the boundary conditions: in the
frame of the perturber, $f$ must approach a Maxwellian distribution with
mean velocity $-{\bf v_g}$ and velocity dispersion $\sigma_c$ at infinity.
Finally, the density is $\rho = \int f d^3 {\bf v}$.  Note that, given the
assumption of a uniform background medium, this treatment is exact, and it
works equally well for a point mass or an extended body.

To linear order (that is, neglecting the small fraction of background
particles that are strongly deflected by the perturber), the wake due
to a point mass can be treated analytically (\cite{mulder}).  In the
frame of the perturber, the density enhancement $\Delta =
(\rho-1)/\rho$ is
\bq
\Delta_{\rm CL,pt} ({\bf x},t) \approx \frac{G M_g}{\sigma^2_c} \frac{1}{r} 
\exp \left(-\frac{\mach^2}{2}(1-\cos^2 \theta) \right) 
\left[ 1 - \mbox{erf}\left(\frac{\mach \cos \theta}{\sqrt{2}}\right) \right],
\label{eq:clptwake}
\eq
where $(r,\theta)$ are polar coordinates centered on the perturber's
current location.  (Note that Mulder's derivation contains a sign
error, though his final expression is correct.  We find $\varepsilon
\rightarrow - \varepsilon$ in Mulder's equations [AIV.3] and [AIV.4].)
This approximation can be derived in a completely independent manner
using the theory of linear response (\cite{colpi}).  Note that the
density structure is 
independent of the propertiers of the perturber and the background
medium if we rescale ${\bf \tilde{r}} = {\bf r}/R_0$, where $R_0 = G
M_g/\sigma_c^2$; this is true for an extended perturber as well.

Because of the lack of shocks, wakes in a collisionless medium are smooth
regardless of the velocity of the perturber, with the peak density
enhancement directly behind the perturber.  Increasing the
velocity focuses relatively more of the wake behind the perturber, forming
a center-filled cone-like structure at high velocities (see
\cite{mulder}, Fig. 2).  This is to be contrasted with the well-defined,
edge-filled cones we find below in collisional media.

\subsection{ Collisional (Fluid-like) Medium }

We calculate the wake in a highly collisional, or fluid, medium using
linear perturbation theory.  The passage of the object perturbs the
background medium, yielding a density field $\rho = \rho_c (1 + \Delta_{\rm
F})$; we assume that $\Delta_{\rm F} \ll 1$.  The first-order continuity
equation is then (see also \cite{ostriker}) \bq \frac{1}{\sigma_c}
\frac{\partial \Delta_{\rm F}}{\partial t} + \nabla \cdot {\bf v_1} = 0,
\label{eq:cont}
\eq
where ${\bf v_1}$ is the perturbation to the velocity field, and the
Euler equation is
\bq
\frac{1}{\sigma_c} \frac{\partial {\bf v_1}}{\partial t} + \nabla
\Delta_{\rm F} = -\frac{1}{\sigma_c^2} \nabla \Phi_{\rm ext},
\label{eq:euler}
\eq
where $\Phi_{\rm ext}$ is the gravitational potential of the
perturbing body.
Combining these equations with Poisson's equation for the potential,
eliminating ${\bf v_1}$, and Fourier transforming over both space and
time yields
\bq
\left( k^2 - \frac{\omega^2}{\sigma_c^2} \right) \hat{\Delta}_{\rm F} =
\frac{4 \pi \mbox{G}}{\sigma_c^2} \hat{\rho}_g.
\label{eq:conteul}
\eq
Here $\rho_g = \rho_g(R,z-v_g t)$ is the specified density field of the
perturber and a hat denotes a Fourier transformed quantity.  
This may be simplified by noting that $\hat{\rho}_g = (2 \pi)^{1/2}
\delta(\omega-k_z v_g)\hat{\rho}_0$, where $\rho_0$ is the density
distribution of the perturber held stationary at the origin.  Taking the
inverse Fourier transform and evaluating the Dirac delta function, we find
\bq \Delta_{\rm F}({\bf x},t) = \frac{4 \pi \mbox{G}}{\sigma_c^2}
\int d^3{\bf k} 
\frac{\hat{\rho}_0}{k^2- \mach^2 k_z^2} e^{i {\bf k} \cdot {\bf s}},
\label{eq:alphamid}
\eq where ${\bf s} = {\bf x} - v_g t {\bf \hat{z}}$.  The integral on the
right may be inverted with the convolution theorem.  However, the inverse
Fourier transform of $\hat{f}({\bf k})=1/(k^2-\mach^2 k_z^2)$ is not
trivial if $\mach > 1$, because in that case the integrand has poles at
$k_z = \pm (k_x^2+k_y^2)^{1/2}/\beta$, where $\beta^2 = \mach^2-1$.
We therefore evaluate the 
integral by transforming to the complex plane and choosing a contour
slightly above the real axis.  If $s_z > 0$, the point of interest is ahead
of the perturber in real space, and causality dictates that the medium
remains unperturbed.  We close the contour at $+i \infty$ in this
case.  Conversely, if $s_z < 0$, we close the contour at $-i \infty$.
Then, using the convolution theorem, 
\bq \Delta_{\rm F}({\bf x},t) =
\frac{G}{\sigma_c^2} \int d^3 {\bf r'} \frac{\xi' \rho_0({\bf
r'})}{\left[(s_z-z')^2 - \beta^2 (R^2 + {R'}^2 - 2 R R'
\cos{\theta'})\right]^{1/2}}, 
\eq 
where
\bq \xi' = \left\{
\begin{array}[l]{ll}
2 & \mach > 1, (s_z-z') + \beta \sqrt{(R^2 + {R'}^2 - 2 R R'
\cos{\theta'})} > 0,\\
1 & \mach < 1, \\
0 & \mbox{otherwise.}
\end{array}
\right.
\label{eq:xidefncom}
\eq
In the case of a point mass, $\rho_0 = M_g \delta({\bf r'})$, this
simplifies to the result found by Ostriker (1999):
\bq
\Delta_{\rm F,pt}({\bf x},t) = \frac{ \xi G
M_g/\sigma_c^2}{\sqrt{(z-v_g t)^2 - \beta^2 R^2}}, 
\label{eq:siptwake}
\eq
where
\bq
\xi = \left\{
\begin{array}[l]{ll}
2 & \mach > 1, (z-v_g t) < -\beta R, \\
1 & \mach < 1, \\
0 & \mbox{otherwise.}
\end{array}
\right.
\label{eq:xidefn}
\eq
As in the collisionless case, the wake structure is independent of the
properties of the perturber and the background medium if we rescale
${\bf \tilde{s}} = {\bf s}/R_0$.  Note that these formulae are not
valid in the case $\mach=1$. 
As shown by Ostriker (1999), the dynamical friction
force due to the wake increases rapidly as the velocity approaches the
sound speed, so such a perturber quickly slows to subsonic speeds.

In the collisional case, the structure of the wake depends strongly on
whether the motion is subsonic or supersonic.  In the subsonic case,
the wake has front/back symmetry relative to the perturber and varies
smoothly.  However, in the supersonic case, the wake forms a Mach cone
trailing the perturber with opening angle $\cot^{-1} \beta$.  The overdensity
at the edges of the Mach cone is infinite for a point mass.  An extended
perturber smooths the singularity, spreading the mass contained in the edge
over a region of width $\sim 2 R_g$.  

For a point mass wake, the column density $\delta \Sigma$ along any
sightline intersecting the cone can be computed analytically and is
independent of the location in the wake,
\bq 
\delta\Sigma = \frac{2 \pi
\rho_c R_0}{\sin i (\beta^2 - \cot^2 i)^{1/2}} \qquad \qquad 
(\cot^{-1} \beta < i < \pi - \cot^{-1} \beta),
\label{eq:ptwakecol}
\eq where $i$ is the angle between the direction of motion of the perturber
and the line of sight.  In the case of an extended perturber, the above
formula is very close to the true column density except near the edges of
the cone, where the column density drops to zero over a width $\sim 2 R_g$.

The singular edge to the cone in the point mass case, and the concentration
of mass in this region for an extended mass may raise some concern about
the linear perturbation analysis used here.  Simulations have shown that,
even for a point mass, the resulting dynamical friction force is
nevertheless quite accurate (\cite{sanchez}), so we expect the linear
treatment to work adequately.  We do find that $\Delta_{\rm F} < 1$
throughout most of the volume of interest (far from the perturber), so
linear theory should be accurate in these regions.  We include a more
in-depth discussion of existing simulations in \S 4.3.

\subsection{ Hydrodynamic Wake }

For a collisional medium, we must add the hydrodynamic disturbance
introduced by the passage of the perturber to the gravitational wake.
If the bound halo is ``stiff,'' it deflects ICM particles and sends pressure
perturbations into the ambient medium.  The hard surface approximation is
expected to be adequate for the ICM gas, because the Larmor radius for
typical cluster magnetic fields is very small ($\ll 1 \pc$), although
particles may still penetrate the perturber if the field lines
intersect the body. The width of the edge of the dark matter halo is
expected to be of order the dark matter particle mean free path.

The density disturbance due to the surface effect may be calculated
analytically far from the perturber if the perturbing body is streamlined
(\cite{landau}, \S 115).  We switch to the rest frame of the perturber and
use coordinates ${\bf s}$ with the origin at the center of the body.  In
this frame, the gas velocity is ${\bf v_0} = -v_0 {\bf \hat{s}_z} + {\bf
v_1}$ and the density is $\rho = \rho_c (1+\Delta_{\rm HD})$, where
$\Delta_{\rm HD}$ is the density perturbation due to this hydrodynamic
effect.  Far from the perturbing galaxy, the gas is only slightly
disturbed and linear theory applies.  If we assume that ${\bf v_1}$ is
related to a velocity potential
$\phi$ via ${\bf v_1} = \nabla \phi$, the Euler equation may be written as
\bq \Delta_{\rm HD} = - \frac{v_0}{\sigma_c^2} \frac{\partial
\phi}{\partial s_z},
\label{eq:denhd}
\eq
where we have assumed that $\phi$ and $\Delta_{\rm HD}$ vanish at
infinity.  Using the continuity equation to eliminate the density
enhancement, we find 
\bq
\frac{ \partial^2 \phi}{\partial s_x^2} +
\frac{ \partial^2 \phi}{\partial s_y^2} -
\beta^2 \frac{ \partial^2 \phi}{\partial s_z^2} = 0.
\label{eq:eulerhd}
\eq
The solution is (\cite{landau})
\bq
\phi = - \frac{v_0}{2 \pi} \int_C^{\ell} \frac{S'(u)\, du}
{\sqrt{(s_z-u)^2 - \beta^2 R^2}},
\label{eq:velpot}
\eq
where $C = \mbox{ max}(s_z+\beta R, - \ell)$, $2 \ell$ is the
length of the perturbing body in the direction of motion, $S(z)$ is the
cross-section of the perturbing body, and a prime represents
differentiation with respect to $u$.  The length $\ell$ of the body is
not well-determined in our analysis; the linear theory is strictly
only valid for bodies with very small opening angles.
We let $\ell = x R_g$, where $R_g$ is the transverse radius of the
object (see \S 3).  Note, however, that the assumption of a
streamlined perturber is not likely to be valid in realistic
circumstances.

In this limit, the hydrodynamic wake forms a cone with a structure similar
to the gravitational wake.  Both are ``edge-filled'' cones with opening
angle $\cot^{-1} \beta$.  In the approximation used above, weak
discontinuities appear at the Mach cones originating from the leading and
trailing edges of the body; in reality, far from the perturber, these are
weak shocks (\cite{landau}; \S 114).  The ICM gas is compressed at the
leading shock and then becomes rarefied as it deflects around the body.
The trailing shock discontinuously increases the gas density.  As equation
(\ref{eq:velpot}) shows, the density distribution of the wake depends on
the cross-section of the body, which is not well-determined by our
treatment (it depends on the detailed initial mass distribution, gas
replenishment in the galaxy, and the nature of the ram-pressure stripping).
Unfortunately, the dependence on the cross-section is substantial,
particularly between the two discontinuities.  In addition, real halos are
not streamlined (\cite{a2142}; \cite{a3667}), but rather have blunt
surfaces.  In this case the leading shock is ``detached'' from the front of
the halo.  The bow shock is normal to the direction of motion on the
collision axis and asymptotically approaches the Mach cone at large
distances (\cite{schreier}).  Within this region, the hydrodynamic motions
are quite complex, but some general conclusions can be drawn.

Behind the trailing shock, the density enhancement due to the
perturbing body falls off rapidly.  This is best illustrated by
evaluating equation (\ref{eq:velpot}) for $R \ll -s_z$.  We find 
\bq
\Delta_{\rm HD} \approx \eta x \mach^2 \left(
\frac{R_g}{|s_z|} \right)^3,
\label{eq:hdwaker0}
\eq
where $\eta \sim 1$ depends on the cross-section of the perturber.
We find that $\Delta_{\rm HD} \ll 1$ throughout most of the cone
interior.  However, the compressions and rarefactions are strongest in
the region bounded by the two shocks discussed above.  We find that in
this region
\bq
\Delta_{\rm HD} \approx \frac{\mach^2}{x^2}
\sqrt{\frac{-s_z - \beta R + x R_g}{2 \beta R}} \, \, p\left(\frac{-s_z -
\beta R + x R_g}{2 x R_g} \right) \qquad
-xR_g < s_z + \beta R < xR_g.
\label{eq:hdwakeedge}
\eq Here $p(y)$ is a function that depends on the precise form of the
perturber's cross-section; in general, it changes sign as $R$ varies.
Because $s_z + \beta R$ is bounded, in this region $\delta \Sigma_{\rm HD}
\propto \Delta_{\rm HD} \propto |s_z|^{-1/2}$.  We find for some generic
cross-sections that $|\Delta_{\rm HD}| > 1$ within the region of interest,
indicating that linear theory is not valid.  Therefore we cannot
meaningfully compare the calculated amplitudes of the variations in this
region to the gravitational wake without detailed simulations of the
interaction process.  However, note that the integrated mass perturbation
along any line of sight passing near the collision axis $\delta \Sigma_{\rm
HD}$ vanishes.  As the body moves, it produces finite outgoing cylindrical
sound waves.  The total displaced mass through any line of sight must
therefore vanish (\cite{landau}, \S 95).

There is yet another effect that can create a wake in a collisional
medium.  Stevens, Acreman, \& Ponman (1999) simulated the effects of
ram pressure stripping on the interstellar medium of
galaxies moving through the ICM.  Most of the gas is stripped into a
plume trailing the galaxy, but they also found ``bow shock'' wakes with
similar structure to those produced by point masses.  The simulation box
size was comparable to the size of the dark matter distribution of the
galaxy, and so it is not possible to infer directly the effects of the
stripped gas on large scales ($\gg R_{\rm gas}$) which are of interest
here.  The finite 
mass of gas in the galaxy (particularly after it has traveled some
distance through the cluster) indicates that the stripped gas should not
have an appreciable effect on the large-scale wake structure (see \S 3.2
below).  

In summary, we expect that hydrodynamic interactions around the
perturber will induce a wake of similar structure to the gravitational
wake in a collisional medium.  Although the overdensity in the interior
of the hydrodynamic wake falls off quickly with distance from the
perturber, the mass between the shocks produced by the leading and
trailing edges of the pertuber decreases much more slowly and, in
fact, linear theory breaks down for a large region around
the perturber.  Therefore we cannot put 
meaningful constraints on the observability of the hydrodynamic wake.  

\section{ Halo Truncation in Galaxy Clusters }

\subsection{ Dark Matter Stripping Processes }

Next we describe how to apply the above formalism to galaxies and groups
moving through X-ray clusters.  We assume that the perturber moves through
a constant density cluster core, surrounded by a softened isothermal
envelope \bq n(r) = \left\{
\begin{array}[l]{ll}
n_c & r < r_c, \\
{2 n_c}/[{1 + (r/r_c)^2}] & r \ge r_c, \\
\end{array}
\right.
\label{eq:clusterden}
\eq where $n_c$ is the total number density of ions and electrons
in the core and $r_c$ is the core radius.  For simplicity, we assume
the dark matter distribution in the perturber to be a singular
isothermal sphere with a specified velocity dispersion $\sigma_g$, so
that $\rho_g = \sigma_g^2/(2 \pi G r^2)$.  The mass contained within a
radius $R_g$ is then $M_g = 2 \sigma_g^2 R_g/G$.  

The radius of the halo is determined by the nature of the dark matter.  If
the dark matter is collisionless, only gravitational variations within the
cluster can affect the halo.  Thus tidal stripping determines the final
size, and the truncation radius may be found by solving
$\bar{\rho}_g(<R_g) = \rho_c$, where $\bar{\rho}_g(<R_g)$ is the mean
density inside $R_g$ (\cite{babul}), as also implied by numerical
simulations (\cite{merritt}).  If the dark matter is strongly
self-interacting, the halo is subject to ram pressure stripping as well.
The truncation radius is then determined approximately by balancing the ram
pressure and the external pressure, $\rho_c v_g^2 = \rho_g(R_g)
\sigma_g^2$.  This simple prescription ignores two competing effects.
First, the supersonic motion of the perturber creates a bow shock leading
it (see \S 2.3), which shields the halo from the ICM ram
pressure.  However, simulations show that for mildly supersonic motion the
bow shock provides relatively little protection (\cite{stevens};
\cite{quilis}).  Instead, recent high-resolution simulations of galaxy
motion through the ICM show that turbulent and viscous stripping are as
effective as ram pressure in truncating gas disks (\cite{quilis}).  More
realistic prescriptions will therefore only increase the contrast between
the FDM and CDM model predictions.  Solving for $R_g$ in the two cases
yields \bq R_g = 164 \left( \frac{5 \times 10^{-3} \cmden}{n_c}
\frac{f_g}{0.2} \frac{0.6}{\mu} \right)^{1/2} \left( \frac{\sigma_g}{300
\kms} \right) \alpha_{\rm DM} \kpc,
\label{eq:rg}
\eq where 
\bq \alpha = \left\{
\begin{array}[l]{ll}
({\sqrt{3} \mach})^{-1} ({\sigma_g}/{\sigma_c}), & 
\mbox{collisional medium,} \\
1, & \mbox{collisionless medium,} \\
\end{array}
\right.
\label{eq:deltasi}
\eq $f_g$ is the gas-to-dark matter mass ratio in the cluster, $\mu$ is the
mean atomic weight per particle, and the subscript ``DM'' refers to the
dark matter medium.  The length scale $R_0$ is \bq R_0 = 30 \left(\frac{5
\times 10^{-3} \cmden}{n_c} \frac{f_g}{0.2} \frac{0.6}{\mu} \right)^{1/2}
\left( \frac{\sigma_g/\sigma_c}{0.3} \right)^2 \left( \frac{\sigma_g}{300
\kms} \right) \alpha_{\rm DM} \kpc.
\label{eq:r0num}
\eq Note that
the scaled size of the perturber is \bq \tilde{R}_g \equiv \frac{R_g}{R_0}
= \frac{1}{2} \left( \frac{\sigma_c}{\sigma_g} \right)^2.
\label{eq:rgscale}
\eq

Studies of galaxies in nearby clusters, particularly Virgo and Coma,
have shown convincing evidence for ram pressure stripping of the
interstellar atomic gas (\cite{warmels}; \cite{cayatte}).  Magri et
al. (1988) showed that the HI flux from spiral galaxies increases with
distance from the cluster center and decreases with the X-ray
luminosity of the host cluster, as one would expect for ram pressure
stripping.  Most recently, Bravo-Alfaro et al. (2000) 
examined 19 bright spiral galaxies in the Coma cluster and found a
clear correlation between depth in the cluster and HI deficiency
relative to isolated spirals.  Rubin, Waterman, \& Kenney (1999) reach
a similar conclusion for the ionized gas disks of spirals in the Virgo
cluster.  

It is much less clear whether dark matter may be stripped in these
environments.  Despite early evidence showing that cluster galaxies tend to
have falling rotation curves (\cite{whitmore}), more recent studies have
not found any systematic difference in the shape of the curves with
location in the cluster (\cite{dale}, and references therein).  However,
these observations do not rule out the FDM model, because the observed
rotation curves are based on H$\alpha$ emission.  This gas is subject
to ram pressure and is therefore at least as compact as the dark matter.  A
better test is to examine the stellar rotation curves, which at these
distances can only be probed with bright tracers such as planetary nebulae
(\cite{ciardullo}; \cite{arnaboldi96}; \cite{arnaboldi98}).  To date, such
studies have been able to examine the inner $\sim 20 \kpc$ of relatively
massive ellipticals, about twice the extent of the emission line rotation
curves.  The observations show no evidence for deviations from the normal
flat rotation curves; however, their limited extent would not exclude a
scenario that is only somewhat less efficient than our FDM model.

Another limit comes from the fact that as the halo loses its
dark matter, luminous matter in the galaxy becomes unbound.  As
a simple model, we consider a spherical galaxy with both the dark
matter and stars distributed as singular isothermal spheres.  In this
case the distribution of stellar velocities is Maxwellian with
standard deviation $\sigma_g$ (\cite{binney}).  We
suppose that the initial radius of stellar matter is $R_i$.  After
stripping, the dark matter radius is reduced to $R_g$; if $R_g < R_i$,
we assume that stars outside of $R_g$ can escape from the galaxy if
their total energy is positive.  The fraction of stars that escape the
galaxy is then 
\bq
f_{\rm esc} = \frac{R_g}{R_i} \int_1^{R_i/R_g} \mbox{erfc}
\left(\sqrt{\frac{2}{r}} \right) dr.
\label{eq:fesc}
\eq
Using the observed relation between the velocity dispersion of a halo
and its effective radius $R_e$ (\cite{merrifield}, p. 206 ), we find
that in the FDM model
\bq
\frac{R_i}{R_g} \approx 1.1 \mach \left( \frac{5 \times 10^{-3}
\cmden}{n_c} \frac{f_g}{0.2} \frac{0.6}{\mu} \right)^{-1/2} 
\left( \frac{\sigma_g}{200 \kms} \right)^{-0.2}
\left( \frac{\sigma_c}{1000 \kms} \right),
\label{eq:rold}
\eq where we have set $R_e = 0.36 R_i$ for a singular isothermal sphere.
The well-known ``fundamental plane'' relates the luminosity (assumed
proportional to the number of stars), the effective radius, and the
velocity dispersion of ellipticals.  The scatter in the relation is only
$\sim 0.07$ dex (\cite{mobasher}).  Because the mass loss mechanism
described above moves galaxies away from the fundamental plane, the scatter
implies that $R_i/R_g \la 1.2$ for galaxies in clusters, or that $\mach \la
1.1$ in typical rich clusters.  However, note that we have neglected the
contribution of the (non-stripped) stars to the mass of the galaxy, which
may be substantial in the inner regions.  This limit is therefore currently
no stronger than that provided by rotation curves of galaxies.

On the other hand, observations of dwarf galaxies in the Local Group
indicate that dark matter stripping may be much more effective than the
CDM model predicts.  At least three dwarf spheroidals (Carina, Draco,
and Ursa Minor) are observed to have stellar tails outside of their
optical cutoff radii (\cite{irwin}).  Significant amounts of dark matter
outside of the optical radii would preclude escape of these stars,
indicating that $R_g \sim 300 \pc$ for these systems (\cite{moore96};
\cite{burk97}).  Our model predicts $R_{g,{\rm CDM}} \sim 5 \kpc$
and $R_{g,{\rm FDM}} \sim 200 \pc$ for a typical dwarf galaxy with
$\sigma_g \sim 10 \kms$, favoring a model that is ``nearly'' FDM.  One
caveat is that, if the orbits of the dwarf galaxies are highly
eccentric, tidal stripping may be much more effective than is inferred
from their present locations.

Heat conduction from the hot ICM to the bound halo dark matter will also
tend to evaporate the halos.  However, the thermal conductivity of a medium
is $\kappa \propto \lambda$, where $\lambda$ is the mean free path of the
particles.  In the extreme FDM limit, $\lambda = 0$ and no heat conduction
occurs; Gnedin \& Ostriker (2000) find that heat conduction may be
neglected over a Hubble time if $\sigma_{\rm DM}/m_{\rm DM} \ga 10^4
\sigm$, where $\sigma_{\rm DM}$ and $m_{\rm DM}$ are the self-interaction
cross-section and mass of the dark matter particles, respectively.  If
$\lambda \neq 0$, heat conduction and ram pressure will work together to
truncate halos, making that process more efficient than we would otherwise
expect.  In the opposite regime of very weakly interacting dark matter
particles, scattering interactions between halo and cluster particles can
also evaporate the halo.  However, in the CDM limit, this process can be
neglected.  Gnedin \& Ostriker (2000) find that the evaporation time is
greater than a Hubble time for $\sigma_{\rm DM}/m_{\rm DM} \la 0.3 \sigm$.
Note that evaporation operates independently of the motion of the galaxy
through the ICM, and it will therefore truncate all subhalos, even those of
central cD galaxies.

\subsection{ Indirect Effects of Dynamical Friction in Clusters }

A common difficulty in all treatments of dynamical friction in a uniform
infinite medium is that the mass contained in the wake diverges with
distance from the perturber.  This divergence is normally avoided by
assuming that the perturber is embedded in a medium of finite extent
(\cite{binney}).  The divergence implies that there must exist a distance
$\tilde{r}_{\rm wake}$ inside of which the wake contains mass $M_g$.
Beyond $\tilde{r}_{\rm wake}$, we expect the second-order effect of the
dark matter wake structure on the observed medium to dominate the effect
induced by the perturber itself.  Using the analytic approximations for the
wake due to a point mass, we find \bq \tilde{r}_{\rm wake} \approx 15
\alpha_{\rm DM}^{-1} \left( \frac{\sigma_g/\sigma_c}{0.3} \right)^{-3}
g(\mach)^{-1/2}.
\label{eq:rwake}
\eq
where
\bq
g(\mach) = \left\{
\begin{array}[l]{ll}
2/(\mach^2-1), & \mbox{FDM,} \\
\int_{-1}^{1}  \exp \left[-\frac{1}{2}\mach^2(1-\mu^2) \right] 
\left[ 1 - \mbox{erf}\left({\mach \mu}/{\sqrt{2}}\right)
\right] d \mu, & \mbox{CDM.} \\
\end{array}
\right.
\label{eq:gmdefn}
\eq 
In the cases we study, we find that $\tilde{r}_{\rm wake}$ is
comparable to or somewhat greater than the box size of our
calculations (and thus larger than cluster cores; see \S 4).
Therefore, we neglect the effect of the dark matter wake on itself and on
the gas wake.  We have confirmed numerically that the dark matter wake has
only a small effect on its own structure within $\tilde{r}_{\rm wake}$.  We
can also examine the extent of the stripped gas wake.  If we let $M_{\rm
gas} = k M_g$ and assume that the stripped gas forms a wake similar
to the one we have calculated, the stripped gas will be contained within a
distance $\tilde{r}_{\rm gas} \sim k^{1/2} \tilde{r}_{\rm wake}\ll
\tilde{r}_{\rm wake}$.  Thus the stripped gas should be confined to the
central region around the perturber.

Note that the dynamical friction force $F_{\rm DF} \propto M_g^2$ in both
the CDM and FDM models, with the coefficient $I_{\rm DF}$ varying by a
factor of a few between the two (\cite{ostriker}).  Because FDM halos are
less massive than CDM halos by a factor of $\alpha_{\rm FDM}$, we expect
that FDM halos will lose energy to dynamical friction at a slower rate than
CDM halos and will therefore sink to the cluster core at a slower rate.
However, FDM halos experience a ram pressure force $F_{\rm ram}$ as well.
In our model, $F_{\rm DF}/F_{\rm ram} \approx 4 I_{\rm DF} /(\mach^4
\tilde{R}_g^2) \ll 1$, so the FDM sink time is determined primarily by ram
pressure.  The different causes of sinking in the two regimes can lead to
observable effects in the galaxy distribution (\cite{moore-coll}).  In
particular, if ram pressure is effective, the time for a galaxy originally
situated a radius $r_b$ from the cluster center to sink to the center is
$t_{\rm ram} \sim 4 r_b/\sigma_c$.  This is generally much smaller than the
sink time for CDM halos; we find \bq \frac{t_{\rm DF,\,CDM}}{t_{\rm
ram,\,FDM}} \sim \frac{4.4 \mach^3}{I_{\rm DF}} \left(
\frac{\sigma_c/\sigma_g}{5} \right)^3.
\label{eq:sinktimes}
\eq
Like Moore et al. (2000), we find that the sink time in the FDM regime
is independent of the properties of the galaxy.  In contrast, the sink time
in the CDM regime has a fairly steep dependence on $\sigma_c/\sigma_g$,
implying that in a given cluster we should observe the massive galaxies to
be more dynamically relaxed than smaller galaxies.  Such mass (or
luminosity) segregation has been observed, although its origins are
controversial because morphology segregation in clusters can mask intrinsic
mass segregation (\cite{kashikawa} and references therein).  Kashikawa et
al. (1998) have found that brighter galaxies are more centrally
concentrated than faint galaxies in Coma, though the degree of
concentration depends on morphology.  Furthermore, Drinkwater, Gregg, \&
Colless (2001) found that the velocity dispersion of the giant galaxies in
the Fornax group is a factor of $\sqrt{2}$ smaller than the velocity
dispersion of the dwarf galaxies, indicating that the former are in a more
relaxed state.  This may be due to the more rapid effects of dynamical
friction on larger galaxies.  A similar effect has been observed for
galaxies in rich clusters (\cite{adami}).

A further observable effect that we have not included in our calculations
is the possibility of a separation between the dark and gaseous components
of the perturber.  In the FDM case, the two components are subject to the
same decelarating forces (ram pressure and dynamical friction), but a CDM
halo feels only dynamical friction.  We therefore expect CDM halos to lead
gaseous halos by an amount that depends on the time history of the
interaction.  Approximate equilibrium can be reached if the gravitational
attraction between the dark matter halo and the gas halo balances the ram
pressure experienced by the gas halo (note, however, that such a separation
may be convectively unstable; \cite{rick-sar}).  We can estimate this
distance if we ignore other forces (including gradients in the cluster
potential); we then find that the equilibrium separation is \bq D_{\rm eq}
= 86 \left( \frac{\gamma}{\mach} \, \frac{ 5 \times 10^{-3} \cmden}{n_c} \,
\frac{0.6}{\mu} \right)^{1/2} \left( \frac{f_g}{0.2} \right) \left(
\frac{\sigma_g/\sigma_c}{0.3} \right)^{1/2} \left( \frac{\sigma_g}{300
\kms} \right) \kpc,
\label{eq:eqsep}
\eq where $\gamma \sim 1$ depends on the structure of the gas and dark
matter halos.  Such a separation has been observed between the galaxies in
the NGC 4839 group and its intragroup gas as the group falls into the Coma
cluster, and it can be attributed to the much smaller ram pressure
experienced by the compact galaxies (\cite{neumann}).  The cluster
parameters of Neumann et al. and the observed values of $\sigma_g/\sigma_c
= 0.3$ and $\mach \approx 1.7$ (\cite{colless}) suggest an equilibrium
separation of $D_{\rm eq} \sim 1 \mpc$, much larger than the observed
separation of $\sim 200 h^{-1} \kpc$, where the Hubble constant $H_0 = 100h
\hunits$.  This is consistent with other suggestions that the merger is
recent, with the NGC 4839 group on its first infall into Coma.
Nevertheless, combining the X-ray map with a weak lensing map in this or a
similar system could constrain the collisional nature of the dark matter
simply by locating the mass peak.

\section{ Results }

Two techniques may be employed to observe the effects of the truncated dark
matter halos.  First, gravitational lensing can constrain the projected
surface mass density of the perturber/cluster system (\S 4.1).
Alternatively, the wake in the ICM gas can be observed through its
X-ray bremsstrahlung emission (\S 4.2).  The former method has the
advantage of directly testing the dark matter profile, but the X-ray
signal is stronger and easier to observe.

In presenting our results, we must ensure that our formalism (in
particular the assumption of a constant density background medium)
remains valid.  If the background medium has $\rho_0 = \rho_0(R)$,
where $R$ is radial distance in the cluster, we have 
\bq
\nabla^2 \Delta_{\rm F} - \frac{1}{\sigma_c^2} \frac{\partial^2
\Delta_{\rm F}}{\partial
t^2} = - \frac{1}{\rho_0} \frac{\partial \rho_0} {\partial R} \left[
\frac{\partial \Delta_{\rm F}}{\partial R} - \frac{1}{\sigma_c^2}
\frac{\partial 
\Phi_{\rm ext}}{\partial R} \right] - \frac{1}{\sigma_c^2} \nabla^2
\Phi_{\rm ext}
\label{eq:rhovary}
\eq
in place of the wave equation with a single source term that led to
equation (\ref{eq:conteul}).  We expect that $(\partial
\Delta_{\rm F}/\partial R) \sim \sigma_c^{-2} (\partial \Phi_{\rm 
ext}/\partial R) \sim 1/R_0$ while $\rho_0^{-1} (\partial
\rho_0/\partial R) \sim 1/r_c$.  Therefore, in order to neglect the
bracketed term on the right-hand side, we require $r_c/R_0 \equiv
\tilde{r}_c \gg 1$.  
A proper treatment in regions with rapidly varying density (as well as
an examination of the effects of shocks and other nonlinearities)
requires numerical simulations.  For these reasons, we discuss our
results in light of existing simulations in \S 4.3. 

\subsection{ Projected Surface Mass Density }

The top two panels of Figure 1 show the surface density of a $\sigma_g =
500 \kms$ group merging with the core of a $\sigma_c=1000 \kms$ cluster
with $n_c = 5.4 \times 10^{-3} \cmden$ and $r_c = 0.4 \mpc$.  The
group has $\mach = 2$.  The top
left panel shows the surface density for the FDM model, and the top
right panel shows the corresponding result for the CDM model.  The
horizontal dashed lines in each panel mark where the Mach cone with
its apex at the center of the perturber intersects the edge of the
cluster core; the apparent suppression of the signal 
below this line occurs only because we do not allow the wake to
propagate out of the cluster core. 

\begin{figure}[htbp]
\includegraphics{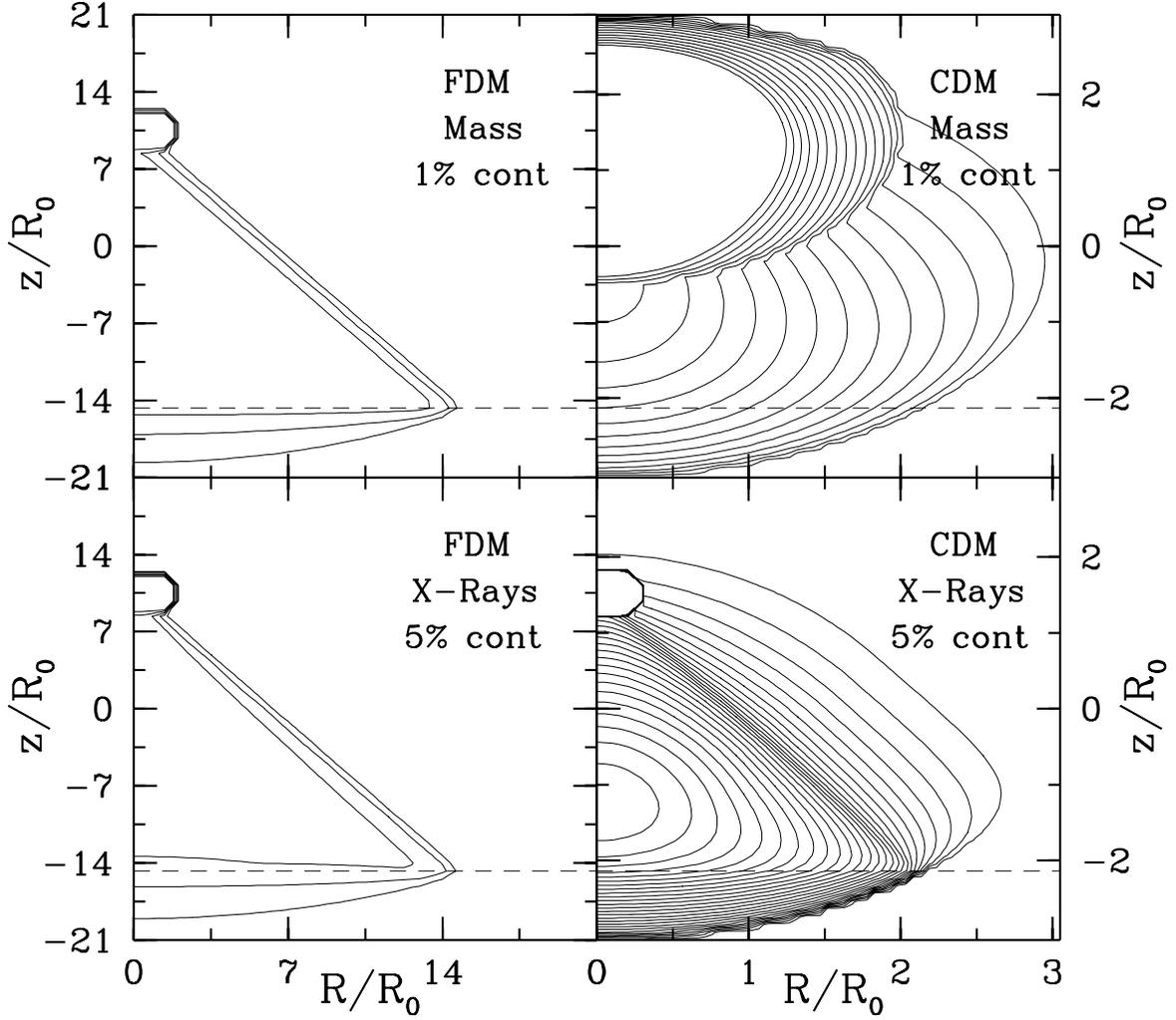}
\vspace{6.2in}
\caption{Results for a $\sigma_g = 500 \kms$ perturber moving with
$\mathcal{M}$ $=2$ through a background cluster with $\sigma_c = 1000
\kms$, $n_c = 5.4 \times 10^{-3} \cmden$ and $r_c = 0.4 \mpc$.  \emph{Top
left:} Projected mass profile, FDM model. \emph{Top right:} Projected
mass profile, CDM model.  \emph{Bottom left:} Surface brightness,
FDM model.  \emph{Bottom right:} Surface brightness, CDM model.
Contours are set at $1\%$ of the peak surface density of the
background cluster in the top panels and at $5\%$ of the peak surface
brightness of the background cluster in the bottom panels.  The merger
occurs in the plane of the sky with the perturber moving along the
$z$-axis.  Axes (in cylindrical coordinates) are in units of $R_0$
appropriate to each panel (see Eq. [\ref{eq:r0num}]).  Horizontal dashed
lines mark the intersection of the Mach cone with the cluster core.}
\end{figure}

The most straightforward way to detect the FDM Mach cone is to look for the
abrupt drop in projected mass at its edge.  The strength of this signal can
be estimated by comparing the (constant) surface density of the cone, given
by equation (\ref{eq:siptwake}), to the surface density of the cluster in
which it is embedded, $\Sigma$.  In the limit in which the impact
parameter of the line of sight is much less than $r_c$, the excess is 
\bq 
\left. \frac{\delta \Sigma}{\Sigma} \right|_{\rm FDM} \sim 
\left( \frac{\pi}{\pi/2 + 1} \right)
\frac{\tilde{r}_c^{-1}}{\sin i \, (\beta^2 -
\cot^2 i)^{1/2}} \qquad \qquad (\cot^{-1} \beta < i < \pi - \cot^{-1}
\beta).
\label{eq:massexcess}
\eq 
Here
\bq
\tilde{r}_c = 13.3 \left( \frac{r_c}{400 \kpc} \right)
\left(\frac{5
\times 10^{-3} \cmden}{n_c} \frac{f_g}{0.2} \frac{0.6}{\mu} \right)^{-1/2}
\left( \frac{\sigma_g/\sigma_c}{0.3} \right)^{-2} \left( \frac{\sigma_g}{300
\kms} \right)^{-1} \alpha_{\rm DM}^{-1}.
\label{eq:rctilde}
\eq
The amplitude depends strongly on the ratio of the velocity
dispersions, so the best candidates are galaxies moving through groups or
groups of galaxies moving through clusters.  The small surface density of
groups makes lensing observations difficult, so we focus on the latter
possibility in this section.

As discussed in \S 2.3, collisional media also give rise to wakes from
hydrodynamic and stripping processes.  The effects of the hydrodynamic wake
well behind the perturber should be small, because $\delta \Sigma_{\rm
HD}$ vanishes to linear order.  However, the stripped dark matter
could enhance the signal immediately behind the galaxy and the leading
bow shock may be quite strong as well; we therefore regard our
calculation as a lower limit to the true enhancement in the FDM case.

Another observable difference between the two pictures is the amplitude of
the mass perturbation induced by gravity.  Tidal stripping in the CDM model
leaves halos much more extended than FDM halos.  They are therefore more
massive and have correspondingly more gravitational influence on the
ambient medium; equation (\ref{eq:massexcess}) shows that $\delta
\Sigma/\Sigma \propto \alpha_{\rm DM}$.  In fact, the enhancement in Figure
1 is somewhat smaller than this because the CDM wake does not have the same
structure as the FDM wake.

A third observable effect is the size of the dark matter halo itself, which
is dramatically different in the two regimes because of the different
stripping mechanisms.  By correlating the mass maps (from lensing) with
X-ray maps, we can compare the relative sizes of the gas halo (which we
know to be stripped by ram pressure) and the dark matter halo.  One method
is that of Natarajan \& Kneib (1997), who analyzed the variation in the
shear of background galaxies in concentric annuli around cluster galaxies
in order to constrain the sizes of their dark matter halos.  So far, the
method has yielded only an upper limit on the sizes of the halos that is
comparable to their tidal radius (\cite{nk-ac114}).  Such a comparison
should be more straightforward in the case of a group of galaxies
merging with a 
larger cluster.  Note that this effect does not necessarily rely on the
extreme FDM model, because a combination of ram pressure, heat conduction,
and evaporation due to scattering will occur across the entire possible
range of parameters for SIDM.

The recent \emph{Chandra} images of sharply defined gas halos in merging
clusters offer excellent opportunities for such comparisons.  In
particular, Markevitch et al. (2000) found two sharp X-ray surface
brightness features in the merging cluster Abell 2142.  They argue that
these edges are formed by ram-pressure stripping during an ongoing merger
in which the two components have passed each other once.  Optical data
indicate that the merger velocity is $\mach \sim 1.5$ and that $\sigma_c
\sim 1200 \kms$ (\cite{oegerle}).  The smaller component has $R_{\rm gas}
\sim 50 h^{-1} \kpc$.  Given the observed parameters of the cluster gas
(\cite{henry}), equation (\ref{eq:rg}) suggests that the velocity
dispersion of the smaller component is $\sigma_g \sim 700 \kms$.  Our FDM
model therefore predicts a mass enhancement $\delta \Sigma/\Sigma \sim
11\%$ behind the smaller subcluster.  In the CDM model, we predict a more
gradual density increase with a peak amplitude $\delta \Sigma/\Sigma \sim
45\%$, although the system is at the limit of the validity of linear
theory.  Brightness edges similar to these have been observed in several
other clusters (\cite{a3667}; \cite{a2256}), indicating that such events
are not uncommon.

An alternate approach to tracing the mass distribution around the subhalo
is to map the distribution of nearby dwarf galaxies (Markevitch \&
Vikhlinin 2001, private communication). Dwarfs behave as collisionless test
particles and so form a wake similar in structure to that shown in the top
right panel of Figure 1.  The amplitude of the variation in surface density
of dwarfs near galaxies and groups can then be used to constrain the mass
of the halo and hence the physics of the dark matter.  Asymmetric
distributions of dwarfs near infalling galaxies have been observed in the
Coma cluster (\cite{conselice98}, 1999).  However, the observed amplitudes
of the density variations are $\delta \Sigma/\Sigma \sim 5-6$, two orders
of magnitude larger than our prediction for a galaxy moving through a rich
cluster, so gravitational wakes are responsible for only a minor part of
this enhancement.  In fact, the observed excess of dwarfs in each system is
contained near or within the tidal radius of the primary galaxy, indicating
that the dwarfs may be bound to the primary or may have even been stripped
from the primary due to ram pressure or strong tidal interactions with
nearby galaxies (\cite{conselice98}).  Achieving observable signal-to-noise
ratios on larger scales will most likely require targeted observations of
the wakes of groups falling into clusters.

For an impact parameter $b$ outside the cluster core, $\Sigma \propto
(1+b^2/r_c^2)^{-1/2}$, while $\delta \Sigma \propto R_0 n(b) \propto
(1+b^2/r_c^2)^{-1/2}$.  Therefore the relative signal is roughly
independent of position throughout the cluster, although the absolute mass
density (and hence the induced convergence and shear in lenses) will
decrease as $b$ increases. The increased number of cluster galaxies outside
the core may make the envelopes more amenable to a statistical analysis.

\subsection{ Surface Brightness Maps }

Even without mass maps, it is possible to use observations of the ICM gas
wake to constrain the dark matter properties.  The ICM gas is best observed
with X-rays, in which the emitted surface brightness is $S = \int
(\epsilon_{ff}/4 \pi) d \ell$.  Here $\epsilon_{ff} \propto \rho_g^2
T^{1/2}$ is the bremsstrahlung (free-free) volume emissivity.  We assume in
the following that the temperature of the gas varies adiabatically with its
density ($T \propto \rho^{2/3}$), because recent studies have shown that
heat conduction is suppressed in the ICM (\cite{vikh-coma}).  Holding the
gas isothermal decreases the signal by a small amount (see below) but does
not significantly affect the structure of the observed feature.

In searching for the wake, one may look for an abrupt drop in surface
brightness at the edge of its cone.  An estimation of the signal strength
in this case is less straightforward analytically, because the surface
brightness of the Mach cone produced by a point mass diverges (due to the
singularity at the edge of the cone).  Fortunately, the singularity is
smoothed out by an extended perturber, and far from the perturber
$\Delta_{\rm gas} \ll 1$
throughout the cone.  In this region, we can expand the $(1+\Delta_{\rm
gas})^n$ that appears in the emissivity and retain only the first-order
term.  Here $n = 7/3$ if the gas temperature varies adiabatically with
density and $n = 2$ if the gas is isothermal.  We then find that 
\bq
\left. \frac{\delta S}{S} \right|_{\rm FDM} \sim 
\left[ \frac{ (\pi/2 + 1) \, n \, (2 n - 1)}{2n - 1 + 2 
\, F(n - \onehalf, n, n + \onehalf, -1)}
\right] \, \left. \frac{\delta \Sigma}{\Sigma} \right|_{\rm FDM},
\label{eq:sbexcess}
\eq
with the same assumptions as in equation (\ref{eq:massexcess}).  Here
$F(x,y,z,p)$ is the Gauss hypergeometric function (see
\cite{gradshteyn}, \S 9.10).  The
numerical factor is $\approx 4$ (or $~5$) if $n=2$ (or $7/3)$.
As with the mass excess, the amplitude of the CDM
wake is enhanced over the FDM wake by a factor $\sim \alpha_{\rm
FDM}^{-1}$.  

Equation (\ref{eq:sbexcess}) shows that the best targets for observations
are the same as for lensing, either galaxies in groups or mergers of rich
groups with clusters.  If the dark matter cross-section increases with
decreasing particle velocity as recently suggested (\cite{hogan};
\cite{wyithe}), galaxy groups may allow us to probe a more
strongly-interacting regime that is closer to the FDM model.

The bottom two panels in Figure 1 show the predicted surface brightness of
the merging group and cluster described in the previous subsection.  The
wake from a halo moving through FDM (or CDM) is shown on the left (or
right).  The horizontal dashed lines again mark the points at which the
central Mach cone intersects the cluster core.  Note that the density
enhancement is in fact greater than unity for $0 < z < 1.5$ and within the
central Mach cone in the CDM case; however, the linear approximation
remains valid in the remainder of this panel and in the left panel.  

Figure 2 shows analogous surface brightness maps for galaxies moving
through a group with $\sigma_c = 500 \kms$, $n_c = 2 \times 10^{-3}
\cmden$, and $r_c = 0.3 \mpc$: the left panels show the wake for a galaxy
similar to the Milky Way, with $\sigma_g = 170 \kms$, and the right panels
show the wake from a typical massive elliptical, with $\sigma_g = 200
\kms$.  Both galaxies have $\mach = 1.5$.  Results for the FDM model are
shown in the top panels while those for the CDM model are shown in the
bottom panels.  Horizontal dashed lines show where the central Mach cone
intersects the edge of the group core.  Note that the maps do not include
emission from the hydrodynamic wake, stripped gas, or gas still bound to
the perturber.  We have also ignored any possible separation between the
dark and gaseous halos (see \S 3.2).

\begin{figure}[htbp]
\includegraphics{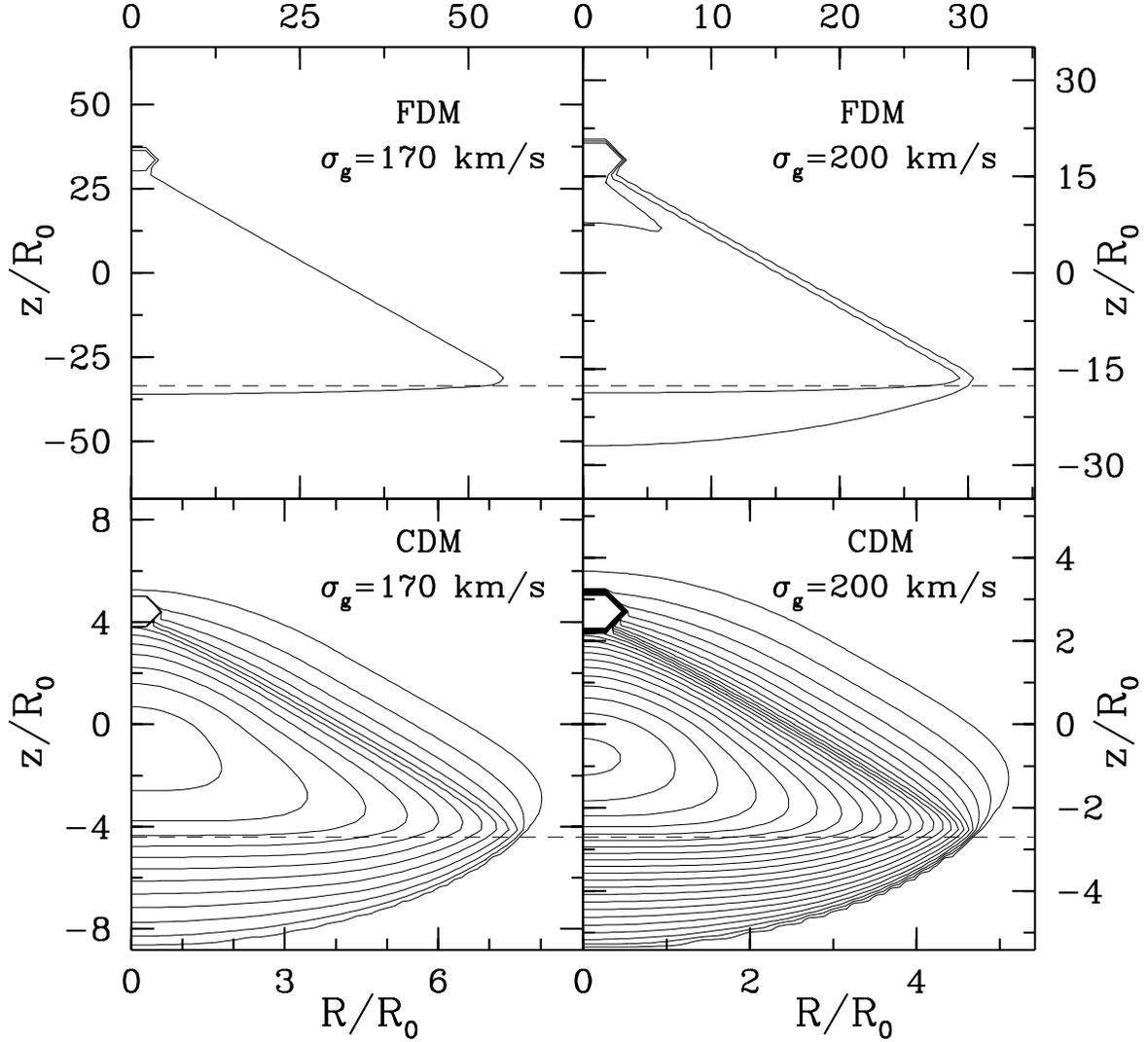}
\vspace{6.2in}\caption{X-ray surface brightness maps for galaxies moving with
$\mathcal{M}$ $=1.5$ through a background cluster with $\sigma_c = 500
\kms$, $n_c = 2 \times 10^{-3} \cmden$ and $r_c = 0.3 \mpc$.  \emph{Top
left:} $\sigma_g = 170 \kms$, FDM model. \emph{Top right:} $\sigma_g = 200
\kms$, FDM model.  \emph{Bottom left:} $\sigma_g = 170 \kms$, CDM model.
\emph{Bottom right:} $\sigma_g = 200 \kms$, CDM model.  Contours are set at
$5\%$ of the peak surface brightness of the background cluster in each
panel.  The merger occurs in the plane of the sky with the perturber
moving along the $z$-axis.  Axes (in cylindrical coordinates) are in units
of $R_0$ appropriate to each panel (see Eq. [\ref{eq:r0num}]).  Horizontal
dashed lines mark the intersection of the Mach cone with the cluster
core.}
\end{figure}

The sharp gradients near the central Mach cone in the CDM panels are not
due to shocks (which we do not include in our formalism) but are
instead due to the assumption of a singular core for the perturber.
In a realistic case, the perturber will have a finite core, which
smooths the gradient over the diameter of the core.  To illustrate
this effect, we show in Figure 3 the wake due to a uniform density
perturber in the FDM (left panel) and CDM (right panel) models.  The
parameters are chosen identically to the right  panels of Figure
2, with the masses normalized to that of the corresponding isothermal
sphere case.  The smoothing of the central gradient by an extended
core should help distinguish CDM wakes from the sharp cones expected
in the FDM model. The morphological differences between the singular
and smoothed cores are much less dramatic in the FDM case because the
halos are truncated at much smaller radii in that case.

\begin{figure}[htbp]
\includegraphics{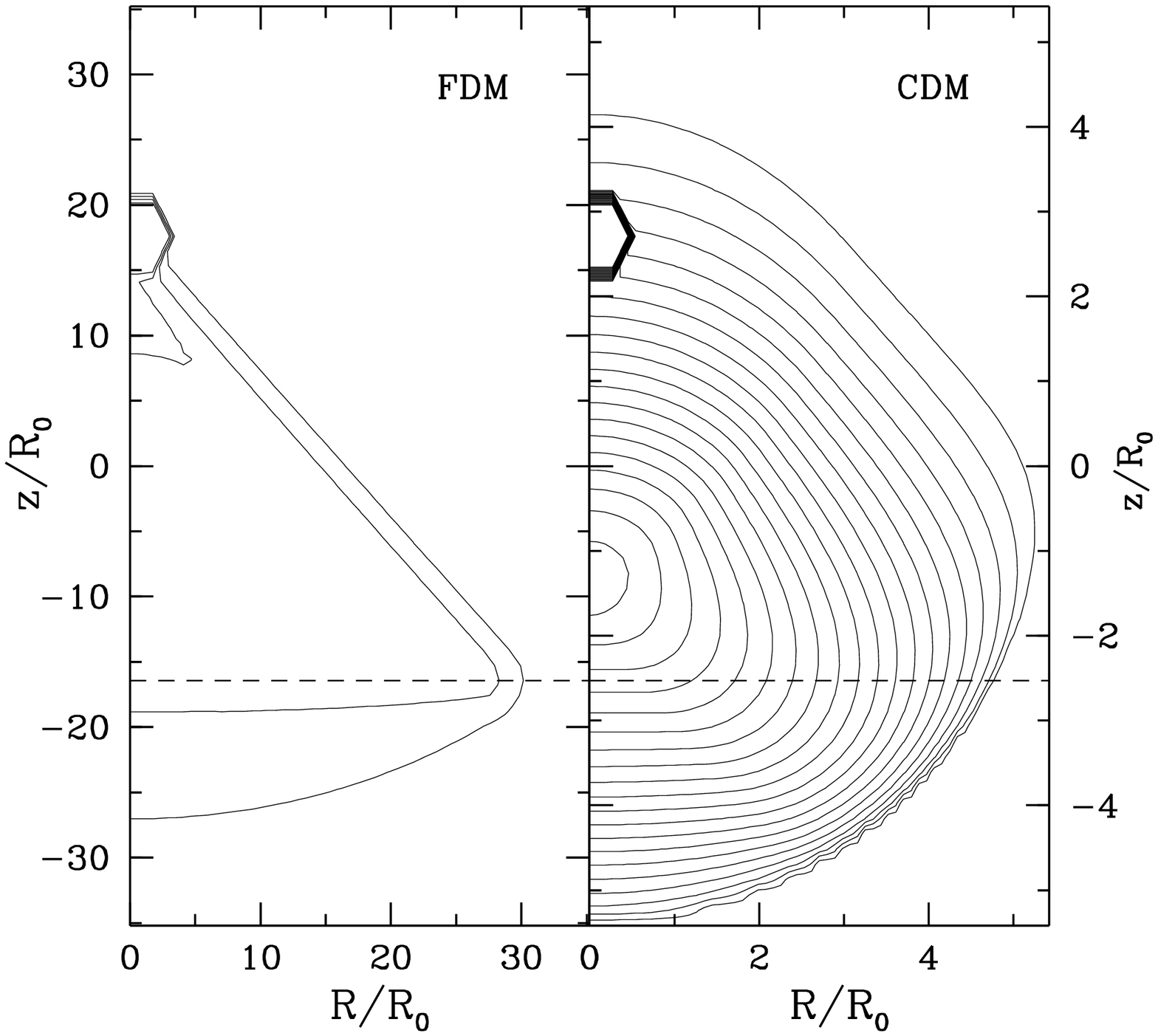}
\vspace{6.2in}\caption{X-ray surface brightness maps for galaxies moving with
$\mathcal{M}$ $=1.5$ through a background cluster with $\sigma_c = 500
\kms$, $n_c = 2 \times 10^{-3} \cmden$ and $r_c = 0.3 \mpc$.  The
perturber is assumed to have a uniform density, with total mass
normalized to a singular isothermal sphere of $\sigma_g = 200 \kms$.
\emph{Left:} FDM model.  \emph{Right:} CDM model.  Contours are set at
$5\%$ of the peak surface brightness of the background cluster.  The
merger occurs in the plane of the sky with the perturber moving along
the $z$-axis.  Axes (in cylindrical coordinates) are in units of $R_0$
appropriate to each panel (see Eq. [\ref{eq:r0num}]).  Horizontal
dashed lines mark the intersection of the Mach cone with the cluster
core.} 
\end{figure}

Limits on the nature of the dark matter from surface brightness maps
are less direct than those from lensing maps,
because we must use the ICM gas wake to infer the size of the dark matter
halo.  The first observational constraint comes from the width of the edge
of the gravitational Mach cone, which is in turn determined by $R_g$.
Let the observable 
size of the gaseous halo be $R_{\rm gas}$. For FDM, we expect that $R_g
\approx R_{\rm gas}$ so that the edge of the cone is tight and follows
closely the observable edge of the perturber.  If, on the other hand, the
dark matter is collisionless, $R_g \approx \alpha_{\rm gas}^{-1} R_{\rm
gas} \gg R_{\rm gas}$, and the edge of the gravitational cone is very
extended.  However, because the ICM gas is
collisional, we expect hydrodynamic and stripping wakes to be present
as well.  The characteristic widths of these wakes is $\sim R_{\rm
gas}$, so we expect a well-defined cone to trail behind the gaseous
halo even in the CDM model.  

Fortunately, there remain signatures in the surface brightness
distribution that can differentiate between the FDM and CDM models.
First, stripped gas can be observationally distinguished from
ICM gas through measurements of metallicity or gas temperature (the
stripped gas will presumably be cooler and more metal-rich than the
ICM gas).  The hydrodynamic wake, like the gravitational wake, is
composed of perturbed ICM gas, which will, if anything, be heated by
compression in the wake.  Therefore we can separate these two wakes
from contamination by the stripped gas.

Even without such a separation, a robust indicator of the CDM model is a
surface brightness enhancement leading the hydrodynamic bow shock.  Because
FDM halos are the same size as the gas halo, causality dictates that their
effects be confined behind the bow shock leading this body.  The 
distance between the bow shock and the front surface of the body may be
estimated if the Mach number and geometry are known (\cite{a3667};
\cite{schreier}).  Any enhancement leading this shock must be due to the
gravitational effect of a CDM halo.  Furthermore, in the CDM model, the gas
halo may trail the dark matter halo because the latter does not experience
ram pressure.  In this case we would observe the gravitational brightness
enhancement to be displaced even farther in front of the hydrodynamic Mach
cone.

Abell 2142 is again a prime candidate for observations.  Our
model predicts a surface brightness excess $\delta S/S \sim 0.4$--$0.5$ in
an FDM cone behind the object and a peak excess $\delta S/S > 1$ in the
CDM model, which should be easily observable in deep observations with
\emph{Chandra}.  Galactic candidates also exist.  NGC 1404, an elliptical
galaxy with $\sigma_g \sim 150 \kms$ (\cite{winsall}) appears to be falling
into the core of the Fornax group, which has $\sigma_c \sim 370 \kms$
(\cite{drinkwater}).  The infall velocity of the system is $\mach \ga 1.1$
(\cite{ferguson}).  Using the X-ray data of Jones et al. (1997), we predict
$\delta S/S \sim 0.12$--$0.15 (r_c/125 \kpc)$ in the FDM model, where $r_c$
is the assumed core radius of the group.  The peak brightness excess in the
CDM model is expected to be $\delta S/S \sim 0.6 (r_c/125 \kpc)$.
Paolillo et al. (2001) 
have made deep \emph{ROSAT} observations of this system, and they see
evidence for a small plume behind the galaxy.  They do not see evidence for
any extended cone emission, although the level we predict is equal to only
one of their contours.  This system therefore provides some weak evidence
against the FDM model.  The morphology is more consistent with a CDM wake,
especially because there appears to be substantial emission leading the
core of the galaxy.  However, higher resolution observations are necessary
in order to precisely determine the ram pressure edge of the galaxy.
Furthermore, the trailing plume is not as extended as one would expect for
the CDM case, although projection effects may disguise the wake.

\subsection{ Comparison to Numerical Simulations }

Numerous groups have simulated cluster mergers (e.g., \cite{roett};
\cite{ricker}; \cite{rick-sar}; \cite{ritchie}) and galaxy/ICM interactions
(e.g., \cite{stevens}; \cite{quilis}) over the past several years.
Unfortunately, neither of these classes of simulations is ideal for
analysis of the gravitational wakes.  In the latter case, the simulation
box size is on the order of the galactic diameter, where nonlinear effects
are strongest and where stripped material dominates the wake.  In the
former case, the mergers studied are generally in the nonlinear regime
(most commonly, two equal mass clusters collide) and in complicated
environments from which it is difficult to isolate the effects of the
gravitational wake.

In fact, our formalism does not apply to existing studies of merging
clusters in the CDM model because the background density varies rapidly.
Such simulations (\cite{roett}; \cite{rick-sar}; \cite{ritchie}) have
$\tilde{r}_c \sim 1$, so our formalism fails in all of these cases, even if
linear theory would otherwise be valid (see equation [\ref{eq:rhovary}]).  In
order to test our model and evaluate the importance of shocks and stripped
gas, we suggest simulations with relatively uniform backgrounds fulfilling
the criterion $\tilde{r}_c \gg 1$.

Recent simulations also show that shocks become significantly less
important as the mass of the perturber decreases because the increase in
the entropy of the core gas is small for such mergers (\cite{ritchie}).
Roettiger, Loken, \& Burns (1997) also find that gradients in the Mach
number of the gas decrease as the mass ratio of the merger decreases,
indicating again that shocks become significantly weaker.  We therefore
expect linear theory to be an adequate approximation in the limit of a
small perturber mass.

To date, there has been little investigation of the direct effects of
cluster mergers in the FDM regime.  Ricker (1998) simulated the merger
of two equally massive gaseous halos, but we do not expect linear
theory to apply in this case.  We therefore cannot use existing
simulations to test the accuracy of our predictions for the FDM model.

A complementary set of simulations to those discussed above have been
performed by S{\' a}nchez-Salcedo \& Brandenburg (1999, 2001), who
investigated the accuracy of the dynamical friction force found by Ostriker
(1999) for a point source traveling through a collisional medium (equation
[\ref{eq:siptwake}]).  They do not directly compare the predicted and
simulated wake structures, but they do find that the drag force expected
from linear theory is accurate to within several percent of the simulation
results, provided that the Coulomb logarithm is chosen appropriately
(\cite{sanchez}).  They have also studied the drag force on a perturber in
a spherically symmetric Plummer model potential (\cite{sanchez2}).  In
their model, $\tilde{r}_c \sim 6$, so we would expect our formalism to be
marginally valid.  Again, they do not analyze the wake structure, but they
do find that the drag force on the perturber is approximately that given by
linear theory if the characteristics of the background medium at the
instantaneous location of the perturber are used.  However, they do not
include the hydrodynamic effects discussed in \S 2.3, so we cannot evaluate
their importance relative to gravitational effects.

\section{ Discussion }

We have calculated the gravitational density wake of a perturber moving
through a uniform collisional or collisionless medium.  We also described
how the dark matter of a galaxy or group moving through a cluster will be
stripped in the cases of fluid-like dark matter (FDM) and collisionless
dark matter (CDM): the former experiences ram pressure stripping, while the
latter undergoes tidal stripping only.  Finally, we calculated the effects
of the wakes in the two different regimes on both the projected cluster
mass and the X-ray surface brightness.

In the FDM model, a supersonic perturber generates a dark matter wake with
a well-defined Mach cone in which the surface density increases
substantially in a relatively narrow region; in the CDM model, the surface
density increases more slowly over a broad region but is of larger
amplitude.  Because the intracluster gas is collisional, we expect the
gravitational wake in the gas to form a Mach cone in both the FDM and CDM
models.  Nevertheless, because FDM halos are truncated at much smaller
radii, the wakes in the two models have significant morphological
differences that can be used to distinguish them.

We found that the signals due to these wakes are potentially observable if
the perturber moves supersonically.  For a large elliptical galaxy
($\sigma_g \sim 250 \kms$) in a rich cluster core ($\sigma_c \sim 1000
\kms$, $n_c \sim 5 \times 10^{-3} \cmden$, $r_c \sim 0.4 \mpc$), the
variation in the surface density is $\delta \Sigma/\Sigma \sim 0.7\%$
across the edge of the FDM Mach cone, while the variation in surface
brightness is $\delta S/S \sim 3$--$4\%$ for the FDM model.  In the CDM
model, we expect a more diffuse wake with signal strength $\sim 6$ times
those in the FDM model.  Fortunately, both signals increase rapidly with
the relative velocity dispersions of the perturber and cluster (see
equation [\ref{eq:massexcess}]).  Therefore, the most favorable systems for
observations have a large $\sigma_g/\sigma_c$: either a gas-rich group
moving through a cluster or a galaxy moving in a group.  For example, the
expected signals for a typical galaxy ($\sigma_g \sim 170 \kms$) in a
galaxy group ($\sigma_c \sim 500 \kms$, $n_c \sim 2 \times 10^{-3} \cmden$,
$r_c \sim 0.3 \mpc$) are $\delta \Sigma/\Sigma \sim 1.3\%$ and $\delta S/S
\sim 7\%$ in the FDM model (or $\delta \Sigma/\Sigma \sim 6\%$ and $\delta
S/S \sim 60\%$ in the CDM model).  Statistical methods may be used to
enhance the signal-to-noise ratio; a related technique, used on
\emph{ROSAT} data to search for small-scale wakes due to ram pressure
stripping, has been developed by Drake et al. (2000).  The relative signal
strength is insensitive to the location within the cluster, so it may be
possible to perform such a statistical analysis over a large volume with
sufficiently deep imaging.

The resolution and sensitivity of the \emph{Chandra} satellite allows
detailed study of the interactions between individual subhalos and the
ICM.  First, 
surface brightness and temperature observations on both sides of the bow
shock leading the perturber (caused by hydrodynamic effects unrelated to
the gravitational wake; see \S 2.3) yield the Mach number of the perturber
through the shock jump conditions (e.g., \cite{landau}).  We can combine
this information with the observed radial velocity of the perturber or with
the opening angle of the shock at large distances to find the inclination
angle of the cone.  When combined with optical observations (yielding
$\sigma_g$ and $\sigma_c$) and X-ray observations (yielding $n_c$, $r_c$,
and gas temperature), we can make an unambiguous prediction of the
enhancement expected from the gravitational wake for the FDM and CDM
models.  This does not include the enhancement due to stripped gas or
deflection of the ICM particles, but for each model it gives a lower limit
to the expected enhancement.  Therefore, observations showing enhancements
smaller than those predicted by the CDM model can be used to falsify that
model.  In addition, temperature maps differentiating the cool stripped gas
from hot ICM gas can help to constrain the cause of any particular observed
enhancement.

Existing X-ray maps of cluster mergers and galaxy motion in groups
have shown no evidence for the sharply defined Mach cones that would
be expected in the FDM model (e.g., \cite{paolillo}); however, there
is very little existing data of a high enough quality to address
this question.  The new generation of X-ray satellites have the
resolution and sensitivity to observe these features with deep
observations.

Our model predicts two systematic variations in the wakes.  First, the
extent of the galaxy halos should increase with distance from the cluster
center: both tidal truncation (for CDM halos) and ram pressure stripping
(for FDM halos) become more effective as the ambient density increases.
We expect $R_g \propto n^{-1/2}$ in both cases, so this systematic
variation cannot be used to test the nature of dark matter, but it can be
used as a consistency check on statistical results for halo sizes.

The second systematic trend can be used to probe the collisional nature of
the dark matter.  Tidal truncation depends only on the relative
gravitational potentials of the perturber and background cluster and is
thus independent of the velocity of the perturber.  Ram pressure, on the
other hand, increases with velocity.  We therefore expect that halo size
will decrease with increasing peculiar velocity relative to the cluster in
the FDM case (where $R_g \propto \mach^{-1}$; see equations [\ref{eq:rg}]
and [\ref{eq:deltasi}]) but not in the CDM case.

Our model for FDM assumes that the dark matter is a perfect fluid,
while many SIDM models treat the dark matter as having a mean-free-path of
order the size of the system.  In such cases, the structure of the wake
(and hence its observability) will lay somewhere between the extreme cases
described here.  Unfortunately, these more moderate cases are not amenable
to analytic treatment.  Instead, one must make use of Monte Carlo
techniques in numerical simulations (e.g., \cite{burkert}; \cite{kochanek})
to analyze the wake structure.  
Such simulations would also help to address some of the other
shortcomings of our model (e.g., the use of linear perturbation theory
in the collisional case,  the assumption of a uniform density
background, the structure of the hydrodynamic wake, and the role of
shocks; see also \S 4.3).

Finally, we outlined in \S 3 a number of consequences of the different
truncation mechanisms acting on halos in clusters, including stellar mass
loss and its effects on rotation curves, the fundamental plane, and local
group dwarfs, relaxation times of galaxies in clusters, and separation of
the dark and gaseous components of galaxy groups.  While indirect, these
effects do not rely on the fluid approximation and may therefore be useful
in constraining the dark matter properties between the fluid and
collisionless limits.

\acknowledgements 

We thank W. Forman, C. Jones, M. Markevitch, and A. Vikhlinin for very
helpful comments on the manuscript.  We would also like to thank P. Ricker
for making unpublished simulation data available to us.  This work was
supported in part by NASA grants NAG 5-7039, 5-7768, and NSF grants
AST-9900877, AST-0071019 for AL.  SRF acknowledges the support of an NSF
graduate fellowship.

\end{document}